# The Ethics of AI Value Chains

Version 3.0
September 18, 2024

Blair Attard-Frost
University of Toronto
Faculty of Information
blair@blairaf.com

David Gray Widder
Cornell University
Cornell Tech, Digital Life Initiative
david.g.widder@gmail.com

**Abstract**
Researchers, practitioners, and policymakers with an interest in AI ethics need more integrative approaches for studying and intervening in AI systems across many contexts and scales of activity. This paper presents AI value chains as an integrative concept that satisfies that need. To more clearly theorize AI value chains and conceptually distinguish them from supply chains, we review theories of value chains and AI value chains from the strategic management, service science, economic geography, industry, government, and applied research literature. We then conduct an integrative review of a sample of 67 sources that cover the ethical concerns implicated in AI value chains. Building upon the findings of our integrative review, we recommend three future directions that researchers, practitioners, and policymakers can take to advance more ethical practices across AI value chains. We urge AI ethics researchers and practitioners to move toward value chain perspectives that situate actors in context, account for the many types of resources involved in co-creating AI systems, and integrate a wider range of ethical concerns across contexts and scales.

**Keywords**
Artificial intelligence, AI ethics, value chains, supply chains, governance, policy

## 1. Introduction

AI ethics is a field of study and practice seeking values, principles, and methods for guiding the implementation of artificial intelligence (AI) systems. Principles and practices of ethical AI often fail to prevent many societal and environmental harms (Attard-Frost, De los Ríos, & Walters, 2023; Greene, Hoffman, & Stark, 2019; Hagendorff, 2020; Lauer, 2021; Morley et al., 2023; Rességuier & Rodrigues, 2020). In response, many researchers have called for AI ethics to be re-centered around new principles or conceptual focal points such as participatory design practices (Birhane et al., 2022a; Bondi et al., 2021), organizational practices (Attard-Frost, De los Ríos, & Walters, 2023; Mäntymäki et al., 2022; Schneider et al., 2023), or relational structures (Birhane, 2021; Crawford, 2021; Crawford & Joler, 2018). Many have also called for new principles and practices of ethical AI based on Ubuntu and Indigenous value systems (Gwagwa, Kazim, & Hilliard, 2022; Lewis et al., 2018; Lewis et al., 2020; Mhlambi, 2020).

In parallel with these developments in AI ethics, policymakers are taking a strong interest in the value chains required to provide resource inputs into and receive resource outputs from AI systems. Emerging regulatory frameworks in the European Union (European Commission, 2018, 2021; European Parliament, 2023), United Kingdom (2023), and Canada (Minister of Innovation, Science and Industry, 2023; Parliament of Canada, 2022) all aim to set obligations on actors within the "supply chain" or "value chain" of AI systems. Meanwhile, research literature has emerged that analyzes the policy implications of "AI supply chains" or "AI value chains" (Brown, 2023; Cobbe, Veale, & Singh, 2023; Engler & Renda, 2022; Kak & West, 2023; Lee, Cooper, & Grimmelmann, 2023; Widder & Nafus, 2022, 2023; Widder & Wong, 2023). However, the emerging regulatory frameworks and policy research literature on AI supply chains/value chains lack a strong theory of AI supply chains/value chains. AI regulatory frameworks and AI researchers often use the terms "supply chain" and "value chain" as though they are interchangeable, when in fact, supply chains and value chains are different types of structures with different ontological, ethical, practical, and policy implications.

In this paper, we present an integrative approach to AI ethics that foregrounds the value chains involved in providing resource inputs to and receiving resource outputs from AI systems. Our study of the *ethics of AI value chains* aims to accomplish two objectives:

> **Research Objective 1** – *Integration of ethical concerns*: We aim to overcome the limitations of many current approaches to AI ethics by integrating a wide range of ethical concerns across many actors, resources, contexts, and scales of activity.

> **Research Objective 2** – *Clarification of value chain implications*: We aim to better theorize and clarify the ontological, ethical, practical, and policy implications of AI value chains.

To accomplish those two objectives, we first review theories of value chains and AI value chains in Section 2. In Section 3, we describe our methodology for an integrative review of literature on the ethical implications of AI value chains. In Section 4, we present the findings of our integrative review. In Section 5, we recommend future directions for researchers and practitioners with an interest in the ethics of AI value chains. We conclude in Section 6 by highlighting the contributions of our review.

## 2. Theory

### *2.1. Value Chains*

In the strategic management literature, the first in-depth theorization of value chains was Porter's (1985) "value chain model." Porter's value chain model specifies five "primary activities" (inbound logistics, outbound logistics, operations, marketing & sales, and service) and four "support activities" (firm infrastructure, human resource management, technology development, and procurement), with each activity transforming resource inputs into valuable outputs and gradually moving resources further downstream in a linear chain-like structure. Later theories in the economic geography literature apply the value chain concept to contexts beyond Porter's predefined "primary" and "support" activities, accounting for the role of value chains in more complex organizational systems and economic networks such as *global value chains* (Gereffi, Humphrey, & Sturgeon, 2006; Humphrey & Schmitz, 2000; Kano, Tsang, & Yeung, 2020) and *global production networks* (Coe, Dicken, & Hess, 2008; Coe & Yeung, 2019; Henderson et al., 2001). More recently, researchers have further extended those theories to critically study the political economies and economic geographies of digital platforms and artificial neural network production (Butollo et al., 2022; Butollo & Schneidemesser, 2022; Ferrari, 2023; Howson et al., 2022a, 2022b).

Researchers in the field of service science, management, engineering, and design (SSMED) have also developed theories of value chains. SSMED researchers conceptualize value chains as linear structures through which value is co-created and progressively added to the chain by a series of actors who exist in diverse service contexts. In these value chains, *value* is conceptualized as phenomenologically co-created preferences for action (Frost, Cheng, & Lyons, 2019), rather than as a positivistic, quantifiable, priceable, and objectively measurable phenomenon as value is generally conceptualized in mainstream economics (Spash 2012). In addition to value chains, SSMED researchers theorize *value networks* as interactive structures that enable value to be co-created between many interdependent actors who are situated across contexts, spaces, times, positionalities, and scales of activity (Edvardsson, Skålén, & Tronvoll, 2015; Frost, Cheng, & Lyons, 2019; Lusch, Vargo, & Tanniru, 2010; Vargo & Lusch, 2016). Foundational to value network ontologies are *resourcing activities*, the activities through which multiple actors across the network assemble and integrate their resources with the goal of co-creating value.

While some regard value *network* ontologies as a conceptually stronger successor to value *chain* ontologies (Basole, 2019; Buhman et al., 2005; Dyer, 2000), others see them as highly compatible. Compatibilist theories view value chains as value network sub-structures through which a set of dyadic actor-actor pairings integrate some of their resourcing activities spatially, temporally, as well as vertically (within a particular industry) and horizontally (across multiple industries) (Alter, 2008; Chen & Chiu, 2015; Lim et al., 2018). For example, Alter's "service value chain framework" assumes that value chains enable linear sets of resourcing activities to be "continuously or repeatedly" (p. 76) performed within pre-negotiated service delivery workflows. Similarly, the "data-value chain" model of Lim et al. characterizes data as a resource from which many networked actors can gradually co-create value through multiple linear chains of data collection, data analysis, and information use activities that are situated across many service contexts. These compatibilist theories of value chains and networks reflect an earlier understanding in the economic geography literature: within network configurations such as global production networks, "there is inevitably an element of linearity or verticality in the structure of its nodes and links" (Coe, Dicken, & Hess, 2008, p. 274). In this view, value chains are linear co-creation structures embedded within larger, non-linear networks of production, distribution, and consumption.

Building upon compatibilist theories of value chains and value networks, we define *value chains* as *co-creation structures that exist within a network of actors and enable patterned resourcing activities to occur between actors*. The resourcing activities that occur within value chains have three main properties:

(1) *Situatedness*: Resourcing activities are situated within specific contexts.

(2) *Pattern*: Resourcing activities are spatially, temporally, and organizationally patterned, and thus capable of recurring with some degree of regularity.

(3) *Value co-creativity*: Resourcing activities are co-created by, perceived differently by, and valued differently by many interdependent actors.

These three properties make the ontologies of value chains markedly different and relatively less linear than the ontologies of supply chains, in which resourcing is typically understood as a series of one-way, downstream movements from producers to consumers.

### *2.2. Supply Chains vs. Value Chains*

*Value chains* have different properties than *supply chains* (Feller, Shunk, & Callarman, 2006). Supply chains are organized according to a "goods-dominant logic," characterized by Vargo and Lusch (2006) as an outdated logic of economic organization in which "tangible output and discrete transactions were central" (p. 4). In contrast, value chains are organized according to a "service-dominant logic" in which "intangibility, exchange processes, and relationships are central" (p. 4). While supply chain ontologies account for a linear set of activities needed to provide tangible resource inputs to production processes (ending in the consumption of those resources), the value chain ontologies of SSMED account for a broader range of intangible and tangible resourcing activities, upstream and downstream relations, and co-creative processes that are simultaneously productive and consumptive.

In the AI ethics literature, the ontological distinction between value chains and supply chains has recently been re-affirmed by Widder and Nafus (2022, 2023). Widder and Nafus call for AI development practices to move away from the task modularity and linearity inherent to supply chain ontologies and toward the broader forms of co-creativity and relationality assumed by value chain ontologies. In the contexts of AI systems, the difference between supply chain and value chain ontologies is crucial: while the supply chains of AI systems scope off a particular set of linear tasks required to make a system usable and make its outputs consumable (thereby ending the supply chain at the "end user" or "consumer"), value chains extend the scope of the system's ethical, practical, and policy considerations into a broader network of co-creative relations. For example, both ontological perspectives can account for the flow of data resources downstream from data subjects to data owners and brokers, model developers, application developers, and end users. However, only value chain ontologies can additionally account for simultaneous upstream flows of financial resources, information resources, and knowledge. A value chain ontology is also more capable of accounting for the production and consumption of material resources, such as the energy and water required to train the model and operate its data infrastructure, or the minerals and fuel required to build and transport the system's hardware components. These materials are omitted from the scope of relatively narrow "AI supply chain", "algorithmic supply chain", and "data supply chain" ontologies that are primarily focused on the downstream flow of data resources (Brown, 2023; Cobbe, Veale, & Singh, 2023; Lee, Cooper, & Grimmelmann, 2023). Figure 1 illustrates this difference in perspective between a hypothetical supply chain ontology of AI systems (focused on vertical production-consumption relations as resources move downstream) and a hypothetical value chain ontology of AI systems (focused on co-creation relations as resources move downstream, upstream, and horizontally through a larger value network).

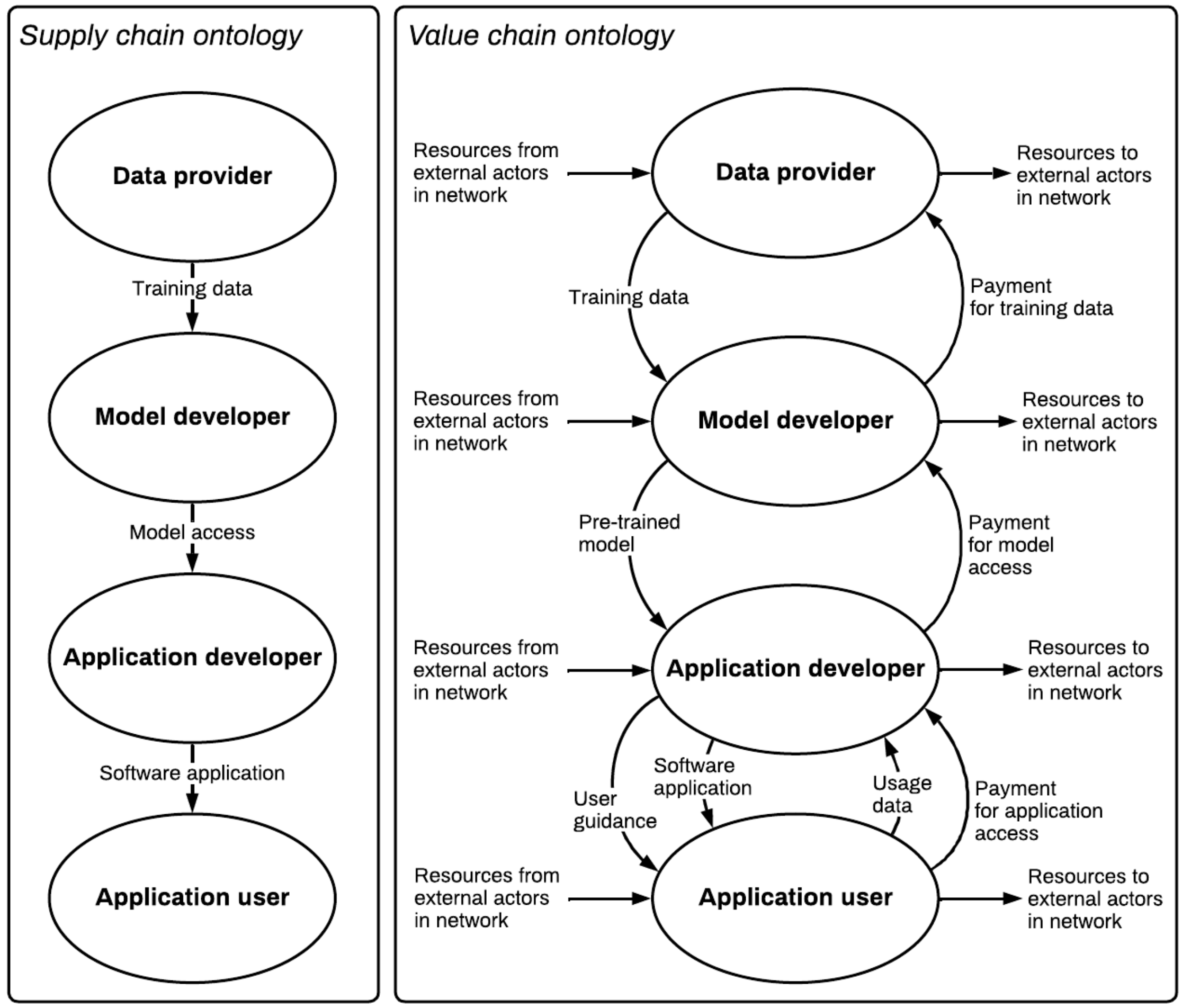

*Figure 1: A supply chain ontology of AI systems contrasted with a value chain ontology of AI systems. The actors and resources that appear here are intended to illustrate key differences between these two perspectives, not to provide a wholly representative or exhaustive view of the actors and resources involved in AI systems.*

## *2.3. AI Value Chains*

A growing body of recent industry, government, and applied research literature examines the value chains of AI systems. Applying our theory of value chains from the previous sub-sections, we define *AI value chains* as *co-creation structures that exist within a network of actors and enable actors to pattern the resource inputs they provide to and the resource outputs they receive from AI systems*.

Much of the applied research literature on AI value chains comes from a strategic management or industrial engineering perspective, examining the role of AI systems in adding value or risk to pre-existing industrial value chains (Chan-Olmsted, 2019; Liu, Chen, & Chen, 2022; Oosthuizen et al., 2020; Staubli, 2022). However, some researchers directly study the ethical and policy implications of providing resource inputs to and/or receiving resource outputs from AI systems. Engler and Renda (2022) propose a typology of AI value chains, common resourcing activities involved in AI value chains, and recommendations for EU policymakers seeking to set more specific obligations for AI value chain participants. Other policy researchers have examined how responsibility and accountability is distributed throughout the value chains involved in supplying data resources to AI systems (Brown, 2023; Cobbe, Veale, & Singh, 2023; Kak & West, 2023; Lee, Cooper, & Grimmelmann, 2023). These studies primarily focus on the software resourcing activities involved in AI systems (e.g., the preparation and use of training and testing data, the purchasing and use of compute, the development and use of models, algorithms, code, and APIs) and propose policy interventions that target those software resourcing activities. Widder and Nafus (2022, 2023) combine theories from computer science and feminist science and technology studies to take a more critical approach to the ontologies and ethics of AI value chains. In examining the practices of 27 AI engineers, they describe AI value chains as "heterogenous, cross-cutting, not always linear social interactions and relations that occupy multiple social locations and cultural logics at the same time" (2022, p. 3). Reflecting SSMED theories of value chains, Widder and Nafus emphasize that AI value chains are situated across social, political, and economic contexts with varying patterns of resource distribution and diverse perceptions of developer responsibility.

Alongside the applied research literature, perspectives on AI value chains from industry represent another emerging body of literature. Similarly to the applied research literature, industry perspectives on AI value chains are predominantly interested in how AI systems can add value to pre-existing industrial value chains by increasing efficiency, effectiveness, or productivity (Appen, 2021; Fife, 2022; Härlin et al., 2023; Shaw & Arkan, 2019). When industry perspectives do discuss the ethics of AI value chains, claims about "responsible AI" or "ethical" AI practices center on the software resources required to develop and use AI systems (e.g., datasets, models, compute, APIs) rather than the social, political, economic, and ecological contexts in which the software resources and resourcing activities are situated.

Many governments take a broader perspective on AI value chains than industry, as governments often aim to intervene in a larger set of societal and environmental impacts than industry is typically concerned with. For example, amendments to the EU's *AI Act* adopted by the European Parliament in June 2023 impose new legal obligations on several value chain actors for conducting data resourcing activities, open-source AI development activities, development and use of "general-purpose AI systems" and "generative foundation models", and environmental impact mitigation activities. Seeking alignment with the EU regulatory framework, amendments to Canada's proposed *Artificial Intelligence and Data Act* (Minister of Innovation, Science and Industry, 2023; Parliament of Canada, 2022) also aim to impose legal obligations throughout Canada's "AI value chain." However, the Canadian framework has not yet set requirements on as broad a range of data resourcing, software resourcing, model development, and environmental impact mitigation activities as the EU framework has. Notably, both the EU and Canadian frameworks neglect to set requirements on the hardware resources involved in developing and using AI systems. This omission indicates that both regulatory frameworks are built upon an incomplete theory of AI value chains that privileges the socio-technical contexts, spatial/temporal/organizational patterns, and value co-creation interactions that are involved in AI *software* lifecycles. The contexts, patterns, and value co-created throughout AI *hardware* lifecycles–along with many other AI value chain actors and resourcing activities–are absent from the incomplete theoretical assumptions underlying these regulatory frameworks.

In addition to software resources and hardware resources, many other types of resources are implicated in AI systems, such as financial resources, knowledge resources, labor and human resources, and governance resources. These resources are input to and output from AI systems by a multitude of actors as they perform various interconnected activities throughout the system lifecycle, such as design, development, deployment, and operation. Ultimately, these actors and their resourcing activities cause many different types of beneficial and harmful impacts to themselves and/or to other actors. Figure 2 illustrates some examples of the actors, resources, activities, and impacts that exist across the value chains of AI systems. The contents of Figure 2 are discussed in more detail as part of our integrative literature review in Section 4.

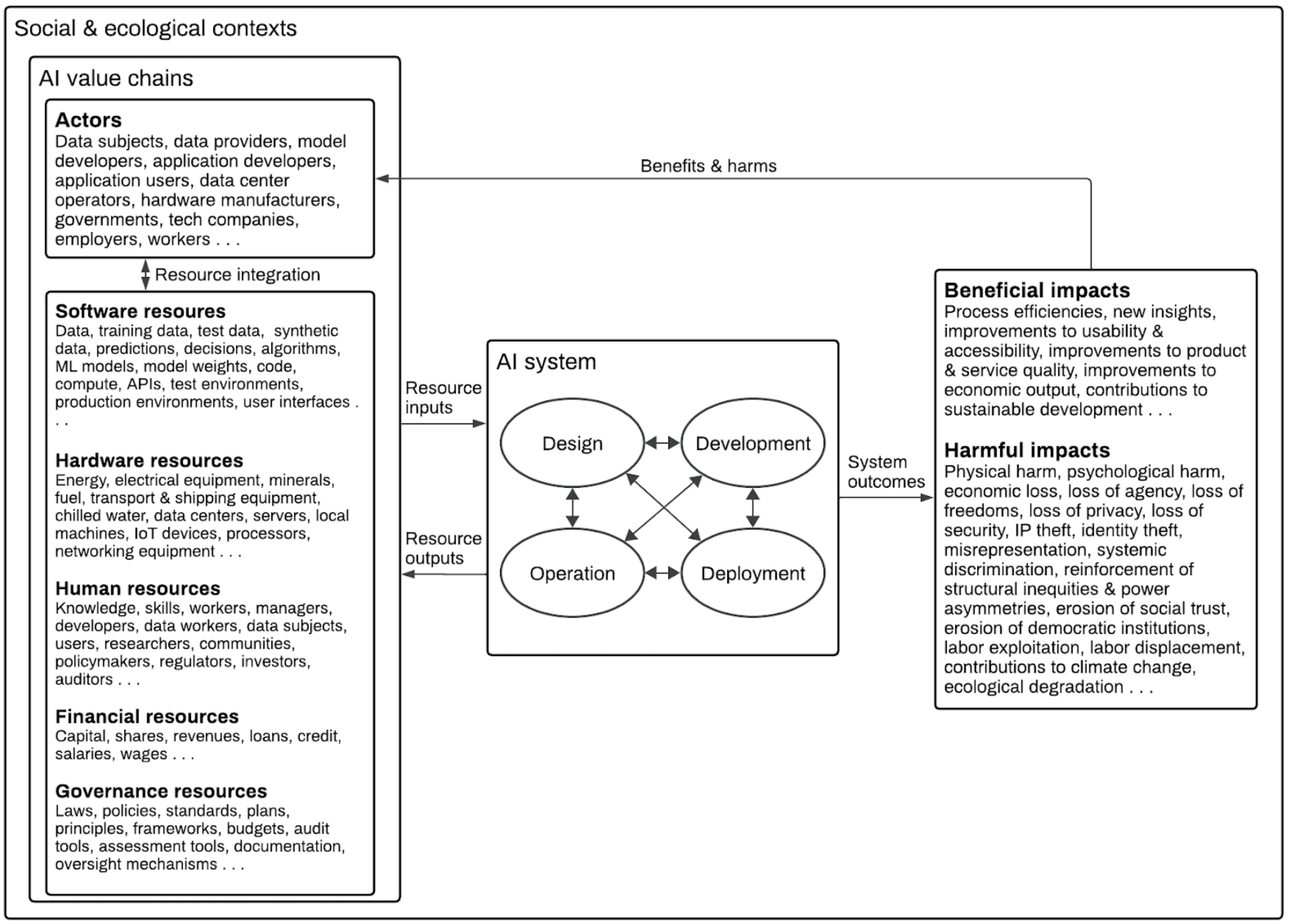

*Figure 2: The process through which many types of actors situated within AI value chains cause beneficial or harmful impacts by providing resources to or receiving resources from an AI system throughout its lifecycle.*

## 3. Methodology

To fulfill our two objectives of integrating AI ethics concerns and clarifying the ontological, ethical, practical, and policy implications of AI value chains, we conducted an integrative review of literature that covers the ethical concerns implicated in our theory of AI value chains. Snyder's (2019) comparison of literature review methodologies recommends integrative review as an ideal method for investigating newly emerging topics to "create initial or preliminary conceptualizations and theoretical models" (p. 336). As a newly emerging topic in need of more detailed conceptual and theoretical development, the ethical implications of AI value chains are a suitable subject for an integrative review. Snyder notes that integrative review "often requires a more creative collection of data, as the purpose is usually not to cover all articles ever published on the topic but rather to combine perspectives and insights from different fields or research traditions" (p. 336). We therefore combined a high-level structure for data collection and analysis with a relatively unstructured set of methods for selecting and integrating the data (see Figure 3). Our goal in making these methodological decisions was to produce a broadly representative and integrative account of the ethical implications of AI value chains, rather than a comprehensive or systematic account of every source that describes the ethical implications of AI value chains. Striving for broad representation and integration of the literature enabled us to more openly explore many newly emerging ideas, relationships, and data sources related to the ethics of AI value chains.

To conduct our integrative review, we first applied the typology of AI ethics concerns developed by Stahl et al. (2022) to create a high-level structure for our data collection and analysis process. This typology covers a uniquely wide breadth of both potential harms and benefits, such as concerns regarding control of data, security and malicious use, concentration of economic power, loss of human autonomy and freedoms, and the development of speculative "superintelligence." Other typologies of AI ethics concerns focus on categorizing harms into fine-grained subsets (Shelby et al., 2023) or on the ethical concerns related to specific types of AI models such as language models or generative AI (Solaiman et al., 2023; Weidinger et al., 2021), while inventories such as the AI Incident Database (2023) and AIAAIC (2023) categorize instances of real-world AI systems causing harm. Applying the typology of Stahl et al. allows us to present our review in a structure that illustrates that our theoretical understanding of AI value chains integrates a wide range of ethical concerns across many actors, resources, contexts, and scales of activity, thereby addressing our first research objective (*Integration of ethical concerns*).

The typology of Stahl et al. includes 4 high-level categories of ethical issues and is further subdivided into 6 types of potential benefits of AI, 8 types and 30 sub-types of potential harms, and 5 "metaphysical issues." For each of those categories and sub-categories of ethical concerns, we identified several value chain actors and resourcing activities related to each of the concerns and recorded those actors and activities in an integrated inventory (see Figure 3 and Appendix A). We then applied a purposive sampling procedure to select sources for inclusion in our integrative review. Purposive sampling requires a high degree of researcher judgment and pre-existing domain knowledge, and is widely used in studies that are intended to provide a rich and illustrative account of some particular phenomena of interest, rather than an exhaustive or statistically representative account of all relevant data (Palinkas et al., 2015; Robinson, 2014). To provide a rich and illustrative account of the ethical issues implicated in AI value chains, we conducted an unstructured series of searches on Scopus, ScienceDirect, and Google using keywords drawn from the ethical concerns, value chain actors, and resourcing activities recorded in our inventory (Appendix A). We selected a source for inclusion in our sample if it met three criteria:

**Criterion 1:** The source was published in a scholarly journal, or by a government organization, researcher, or journalist with expertise in AI ethics.

**Criterion 2:** The source provides a detailed description of one or more of the ethical concerns recorded in our inventory.

**Criterion 3:** The source provides a detailed description of the actors and resourcing activities implicated in those ethical concerns.

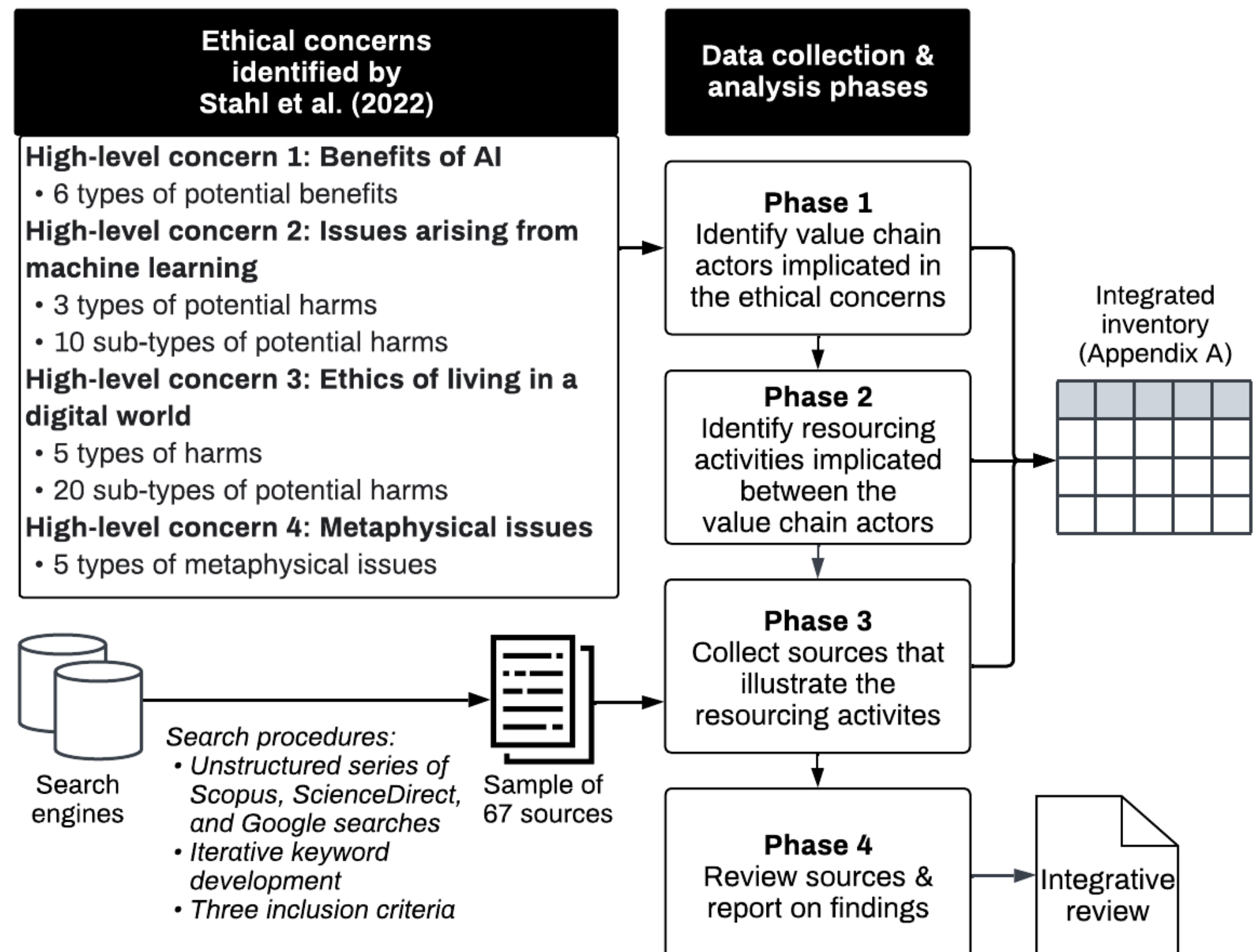

*Figure 3: Diagram of our data collection and analysis process.*

We iteratively developed search keywords using these criteria until we had collected a sample of sources sufficient to illustrate all ethical concerns, actors, and activities in our inventory. We then recorded these sources in our inventory alongside their corresponding issues, actors, and activities (Appendix A). This procedure resulted in 67 sources being included in our sample from research literature, grey literature, and news media that collectively illustrate a wide range of ethical concerns implicated in the AI value chain concepts we theorized in Section 2 (e.g., AI value chains, value chain actors, resources, resourcing activities). We reviewed each source in our sample and integrated our findings into a written description of the ethical concerns, value chain actors, and resourcing activities represented in the sources, thereby addressing our second research objective (*Clarification of value chain implications*). A full inventory of the ethical concerns, value chain actors, resourcing activities, and sources sampled in our integrative review can be found in Appendix A. Our integrative review of the 67 sampled sources is presented in the following section.

## 4. Ethical Implications of AI Value Chains

### *4.1. AI Value Chains & Benefits of AI*

Stahl et al. (2022) note that although AI ethics usually foregrounds the harms of AI systems, AI systems may also present benefits that should be accounted for. The potential benefits include: insights or efficiencies from automating the processing of large volumes of data to make predictions, decisions, or generate synthetic data outputs; improvements in economic output and reductions of environmental damage as a result of more effective and efficient production processes; contribution to United Nations Sustainable Development Goals (SDGs), as well as other international and national pursuits of socially beneficial AI adoption. However, accounting for these potential benefits within a theory of AI value chains enables us to identify many concomitant harms. Novel insights or gains to efficiency in some parts of an AI value chain may raise new risks in other parts of the value chain (Cobbe, Veale, & Singh, 2023; Gansky & McDonald, 2022; Widder & Nafus, 2023). Contributions to SDGs or “AI for good” initiatives may only be successful relative to a narrow set of values and measures (Aula & Bowles, 2023; Madianou, 2021; Moore, 2019). Economic prosperity or environmental benefits may be inequitably distributed across different groups, communities, or geographies. While AI systems may enable some value chain actors to co-create mutually beneficial outcomes, pre-existing structural injustices in the social contexts of AI systems warrant an assumption that the same systems will also result in harmful outcomes for other actors, particularly for those who belong to historically marginalized communities (Birhane, 2021; Hind & Seitz, 2022).

### *4.2. AI Value Chains & Issues Arising from Machine Learning*

Stahl et al. (2022) describe many ethical concerns related to the use of machine learning (ML) technologies and methods in AI systems. These concerns include issues related to (1) *control of data*, (2) *reliability*, and (3) *lack of transparency*.

*Control of data* in AI value chains has been widely studied. The resources and activities required to regulate data–such as public funding, policy development, and enforcement of data protection laws in the training and application of ML models–are a significant concern related to the control of data in AI value chains (European Parliament, 2020; MacKinnon & King, 2022; Veale, Binns, & Edwards, 2018). Many other resourcing activities are implicated within the broader domain of data governance, such as informed consent from data subjects in data collection and data use activities, as well as the sale, purchase, brokerage, and ownership of ML training and testing data (Crain, 2018; Lamdan, 2022). Knowledge and expertise acquired by ML experts is also required to develop ML models, and to identify how vulnerabilities in ML models might be exploited through methods such as inversion attacks or injection attacks (Greshake et al., 2023; Wang et al., 2022). This need for specialized knowledge resources to conduct activities such as ML development and vulnerability testing implicates education and training programs for ML and ML security in AI value chains. Corporate capture of the financial and data resources needed to conduct ML research is also a concern related to control of data (Whittaker, 2021).

Control of data issues can be observed in many real-world cases. For example, the company Clearview AI scraped billions of images from platforms such as Facebook and YouTube to develop facial recognition and surveillance applications of ML that have been used by thousands of law enforcement agencies globally (Hatmaker, 2022; Perrigo, 2022). Clearview's data collection practices raise concerns regarding consent, ownership, regulation, data resourcing, financial resourcing, and other public sector resourcing activities further upstream and downstream from Clearview. Similar ethical concerns are implicated in the value chains of generative AI systems such as ChatGPT, Stable Diffusion, and Midjourney, which are trained on large volumes of data scraped from the open web, typically without explicit consent from the creators or copyright owners of that data (Lee, Cooper, & Grimmelmann, 2023). In generative AI value chains, control and ownership of data is an issue of particular importance to value chain actors such as generative AI developers, AI users, and artists and other creative workers (De Vynck, 2023; GitHub Copilot Litigation, 2023; Stable Diffusion Litigation, 2023; Vincent, 2023).

Activities and ethical concerns related to the *reliability* of ML methods and applications have also been widely studied. Inaccurate predictions, decisions, and other data outputs created through the use of unreliable ML models cause many social, political, economic, physical, and psychological harms (Angwin et al., 2016; Bender et al., 2021; Grote & Berens, 2022; Mökander & Axente, 2023; Rankin et al., 2020). The implementation of effective quality assurance practices for ML model training, testing, and management at multiple points throughout AI value chains is viewed as essential for reliable and ethical ML applications (Burr & Leslie, 2023; Eitel-Porter, 2021). The use of cloudwork platforms and labor outsourcing practices to improve data quality, model reliability, and accuracy can also cause social, political, economic, physical, and psychological harms by subjecting data workers to physically and psychologically unsafe working conditions (Irani, 2015; Miceli & Posada, 2022; Miceli, Posada, & Yang, 2022; Perrigo, 2023).

*Lack of transparency* in ML methods and applications presents major ethical concerns. Transparency of the funding sources for ML research and development is one such concern (Ahmed, Wahed, & Thompson, 2023; Ochigame, 2019; Whittaker, 2021). Documentation, disclosure, and explanation of the data and computational resources, processes, and outcomes involved in ML development activities and automated decision-making activities is another transparency concern (Miceli et al., 2022; Mitchell et al., 2019; Raji et al., 2020). Also of concern is disclosure of the extent to which diverse stakeholder knowledges were included in ML model design, development, and use–particularly the inclusion of vulnerable data subjects and historically marginalized groups (Birhane et al., 2022a, 2022b; Widder & Nafus, 2023). Distribution of accountability for harms amongst AI value chain actors represents another concern, as accountabilities can only be fairly distributed across actors if the resourcing activities are made transparent to one another and to authorities (Bartneck et al., 2020; Brown, 2023; Cobbe, Veale, & Singh, 2023; European Commission, 2022; Zech, 2021). Additionally, practices of collective organizing and resistance against harmful AI systems are an important ethical issue in cases where an ML application has caused harm due to a lack of fairness, accountability, and transparency in its value chain (e.g., ACLU, 2023; Broderick, 2023).

### *4.3. AI Value Chains & Ethics of Living in a Digital World*

Stahl et al. (2022) outline many ethical concerns related to harms caused by AI systems as a consequence of "living in a digital world." These concerns include: (1) *economic issues*, (2) *justice*, (3) *human freedoms*, (4) *broader societal issues*, and (5) *unknown issues*.

Economic issues are especially significant in AI value chains. The use of automation and biometrics in human resources practices–along with related labor resourcing activities such as the hiring, contracting, dismissal, and surveilling of workers–represents a political-economic issue of ethical concern in AI value chains (Bales & Stone, 2020; Hickok & Maslej, 2023), as does distribution of wealth, capital, and other financial resources and labor exploitation across AI value chains (Dyer-Witheford, Kjøsen, & Steinhoff, 2019; Miceli & Posada, 2022; Miceli, Posada, & Yang, 2022). Open-source development of AI systems and open access to data, code, and other software resources also pose concerns related to asymmetries of political and economic power (Langenkamp & Yue, 2022; Masiello & Slater, 2023; Widder, West, & Whittaker, 2023).

One particular case involving many of these economic issues is OpenAI's outsourcing of data labeling activities to workers employed by Sama AI in Kenya, many of whom were psychologically harmed and undercompensated during their employment (Perrigo, 2023). In another case, Amazon's development, use, and subsequent disuse of a hiring automation tool that discriminated against women represents another issue related to harmful labor resourcing activities (Dastin, 2018). The consolidation of models and datasets in an increasingly small group of private sector actors further demonstrates that political and economic harms can be caused by resource distribution imbalances across AI value chains (Ahmed, Wahed, &

Thompson, 2023), as do the summer 2023 labor strikes of the Writers' Guild of America (WGA) and Screen Actors Guild-American Federation of Television and Radio Artists (SAG-AFTRA). WGA and SAG-AFTRA workers demanded that their employers refrain from using their likenesses or union-protected creative materials in generative AI training datasets, as well as from requiring them to use generative AI applications in their work activities without their consent (Broderick, 2023; Webster, 2023).

There are many ethical concerns related to *justice* in AI value chains. For example, public sector procurement, development, and use of automated decision systems in public service and courtroom contexts has caused harm to vulnerable groups (Angwin et al., 2016; Eubanks, 2018; Gans-Combe, 2022; Mulligan & Bamberger, 2019). The inclusion of knowledges and perspectives from historically marginalized groups in AI education, development, and governance processes is another concern in resourcing the knowledge required to ethically develop AI systems (Birhane et al., 2022a; West, Whittaker, & Crawford, 2019). The just distribution of value co-creation outcomes across AI system lifecycles is a concern to many researchers, as is the potential for macro-scale social, political, and economic injustice caused by widespread AI adoption (Dyer-Witheford, Kjøsen, & Steinhoff, 2019; Pasquale, 2020; Solaiman et al., 2023).

Many resourcing activities in AI value chains have impacts on *human freedoms*. Ethical concerns related to human freedoms often overlap with economic issues and justice issues, as these concerns typically stem from a structural lack of freedom to access a particular kind of resource, which in turn, perpetuates a further loss of freedoms. For example, harmful outcomes of exploitative labor outsourcing and algorithmic discrimination often result in marginalized groups experiencing a further loss of access to resources needed to pursue social, political, and economic opportunities (Angwin et al., 2016; Eubanks, 2018; GPAI, 2022; Miceli & Posada, 2022). Restrictive access to public sector and private sector data, information, and computational resources can also result in disproportionate levels of access to and benefit from those resources becoming further reinforced between individuals, groups, sectors, and governments (Ahmed, Wahed, & Thompson, 2023; Whittaker, 2021).

What Stahl et al. (2022) refer to as *broader societal issues* is a category of ethical concerns containing a variety of large-scale impacts on potentials for physical conflict, environmental degradation, and erosion of democratic institutions. For example, military and police procurement of use-of-force and surveillance applications represents a broad societal issue in which many value chains become implicated in potentials for causing physical conflict and violence (Hoijtink & Hardeveld, 2022; Mahoney, 2020; Mulligan & Bamberger, 2019; Taddeo et al., 2021). Social filter bubbles created from the algorithmic profiling and manipulation of social media users is another broad societal issue, implicating many value chains in potentials for causing social, political, physical, and psychological harms (Krönke, 2019; Woolley, 2018). Energy and water are material resources required to train AI models and operate AI systems, and these material resourcing activities also constitute a broad societal issue. Energy and water usage in AI hardware and infrastructure results in substantial carbon emissions, depletion of freshwater reserves, and other global and local environmental harms (GPAI, 2021; Li et al., 2023; Luccioni & Hernandez-Garcia, 2023). In addition to energy and water, mineral extraction and other mining, manufacturing, transportation, and assembly processes involved in the material resourcing of AI systems all represent a broad societal issue, as does the disposal and recycling of environmentally harmful electronic waste at the end of AI hardware lifecycles (Crawford, 2021; Crawford & Joler, 2018).

What Stahl et al. (2022) refer to as *unknown issues* is a category of ethical concerns containing a variety of complex harms that are difficult to predict the potential consequences of. For example, malicious value chain actors might engage in unforeseen misuses of personal data, digital identities, misinformation, or other resources in their malicious development and/or use of AI systems (Brundage et al., 2018). The enforcement of excessively strict or excessively permissive AI regulations may also cause a variety of complex social, political, and economic harms (Ada Lovelace Institute, 2021; Smuha, 2021). Additionally, excessive funding of AI research that prioritizes finding solutions to the wrong problems might result in some AI risks and harms becoming less foreseeable and/or preventable (Tiku, 2023).

### *4.4. AI Value Chains & Metaphysical Issues*

Stahl et al. (2022) describe several “metaphysical issues” pertaining to speculative ethical concerns such as machine consciousness, artificial moral agents, artificial “super-intelligence”, and changes to “human nature” enabled by new AI technologies. These “metaphysical issues” are purely speculative. However, the ethical implications of these hypothetical activities are comparable to the ethical implications of empirically observable resourcing activities. For example, concerns related to the distribution of benefit/harm through the development of a speculative "autonomous" *artificial moral agent* are comparable to real-world concerns related to *human moral agents* distributing benefit/harm through the development and use of automated systems. Similarly, issues of resource distribution, consolidation, and power asymmetry arising from the development of speculative *superintelligent agents* are comparable to issues of resource distribution, resource consolidation, and power asymmetry that exist between real-world human agents.

Some researchers convincingly argue for disregarding these speculative ethical concerns and instead accounting for real, present harms (Gebru & Torres, 2023; Torres, 2023). We advance these arguments by noting that the ethical concerns underlying these speculative "metaphysical issues" are futurological extensions of ethical concerns that can already be observed in real-world, present-day AI value chains. Therefore, greater study can and should be given to the empirically observable AI value chains. We also note that AI value chain theories and methods are flexible enough to account for the benefits and harms of AI systems across multiple spatial, temporal, and organizational scales (including those benefits and harms that exist only in speculative futures not meriting significant study).

## 5. Future Directions for Research, Practice, & Policy

Future research, practice, and policy should more comprehensively account for, integrate, and intervene in the range of ethical concerns, value chain actors, and resourcing activities we outlined in the previous section. An integrative approach to accounting for and intervening in the ethics of AI value chains requires intervening in many types of resources, activities, and societal and environmental impacts. Software resources and the activities performed to integrate software resources into AI systems–for example, the collection and preparation of training and test data, the development and deployment of models, and operation and monitoring of a model throughout its software lifecycle–raise ethical concerns at many upstream and downstream sites in AI value chains. In addition to software resources, many other types of resources, activities, and impacts raise ethical concerns, including the resourcing of hardware, finances, knowledge, labor, and governance mechanisms throughout the lifecycles of AI systems.

There are three opportunities for researchers and practitioners to further investigate the wide range of ethical concerns implicated in AI value chains:

(1) *Conduct more empirical and action research* into the specific ethical concerns, value chain actors, and resourcing activities we outlined in Section 4. Future research agendas could include, for example, empirical and participatory studies of the impacts of generative AI development and use on artists and workers, or studies of the impacts of outsourcing practices on marginalized workers in AI value chains. By collecting and analyzing more quantitative and qualitative data pertaining to a variety of real-world AI value chains and their related actors and activities, researchers can provide a rich evidence base upon which other researchers, practitioners, and policymakers can develop further research, practice, and policy on the basis of. Additionally, by empowering value chain actors to participate in research activities, identifying their concerns and needs, and developing interventions that are designed to satisfy their needs, researchers can generate more detailed insights on stakeholder perspectives, best practices, policy gaps, and policy options.

(2) *Develop and apply theories and methods* for systematically modeling AI value chains, analyzing a diverse range of ethical concerns in those value chains, and enacting interventions in those value chains. Many frameworks for systematically modeling and analyzing value chains and value networks can be applied to studies of AI value chains, such as the service system analysis framework of Frost, Cheng, and Lyons (2019) and the data value chain framework of Lim et al. (2018). These and similar frameworks can help to ground future research on the ethics of AI value chains in stronger value chain theories and methodologies.

(3) *Design and implement ethical sourcing practices* across all of the value chains that provide resource inputs to or receive resource outputs from AI systems. Ethical sourcing practices have been adopted in with different degrees of effectiveness in many industries, and are intended to support in “managing all processes of supplying the firm with required materials and services from a set of suppliers in an ethical and socially responsible manner” (Kim, Colicchia, & Menachof, 2018, p. 1033). In the context of AI practices, ethical sourcing requires all actors that provide resources inputs to or receive resource outputs from AI systems to account for a diverse range of impacts that their activities have on society and the environment (Widder & Wong, 2023). Many frameworks for guiding ethical sourcing practices in AI systems have been developed by academic researchers, such as documentation and auditing frameworks for data and model resourcing (Mitchell et al., 2019; Miceli et al., 2022; Raji et al., 2020). Fairwork's principles and practices for preventing harms to workers across AI value chains (Fairwork, 2023; GPAI, 2022) and Global Partnership on AI's principles and practices for mitigating the harmful environmental impacts of AI systems (GPAI, 2021) represent two ethical sourcing frameworks that cover a range of ethical concerns related to labor and the environment.

Governance mechanisms such as industry standards, certification programs, guidance documents, and codes of conduct should also be used to support the implementation of ethical sourcing practices across AI value chains. Many existing standards, certification programs, guidance documents, and codes of conduct focus on a narrow socio-technical context, while affording minimal or no attention to concerns implicated in the broader social, political, economic, material, and ecological contexts of AI value chains (Government of Canada, 2023a; ISO, 2023; NIST, 2023; Responsible Artificial Intelligence Institute, 2022). In contrast, the Treasury Board of Canada Secretariat has published a voluntary guide for using generative AI applications in Canada's federal public sector that provides principles and best practices for a wide-ranging set of ethical concerns, such as AI literacy development, professional autonomy, and environmental impact mitigation (Government of Canada, 2023b). Future iterations of these and other AI governance mechanisms–such as legislation, regulations, and other policy instruments–should be used to implement more comprehensive principles and best practices for ethical sourcing across the many actors, activities, and contexts of AI value chains.

## 6. Conclusion

We have reviewed and synthesized theories of value chains and AI value chains. We have also conducted an integrative review of recent literature on the ethical implications of AI value chains. We have therefore accomplished our two objectives of (1) integrating AI ethics concerns into our conceptualization of AI value chains, and (2) clarifying the ontological, ethical, practical, and policy implications of AI value chains. In doing so, we have made a scientific contribution to the theoretical development, scholarly knowledge, and practitioner knowledge of AI value chains. The opportunities for future research and practice we outline also represent a significant practical contribution, as these opportunities provide a preliminary agenda for further advancing studies and practices of ethical AI value chain governance.

AI value chains will remain a focal point of AI ethics and governance initiatives into the foreseeable future. As those initiatives continue to unfold, researchers, practitioners, and policymakers with an interest in the ethics of AI value chains can look to this paper for guidance in conceptualizing and conducting their work. AI ethics must advance beyond decontextualized discussions of ethics and toward value chain perspectives that situate actors in context, account for the many types of resources involved in co-creating AI systems, and integrate a wider range of ethical concerns across contexts and scales.

## Appendix A: Integrated Inventory of Ethical Concerns, Value Chains Actors, Resourcing Activities, & Sampled Sources

*High-level ethical concern 1: Benefits of AI*

| Lower-level issues identified by Stahl et al. (2022) | Examples of related value chain actors | Examples of related resourcing activities | Related sources |
|---|---|---|---|
| Novel insights from data, efficiency improvement, economic benefits, environmental benefits, contribution to sustainable development goals, AI for Good | • Industry<br>• Governments<br>• Intergovernmental bodies<br>• Civil society | • Analytics platform development & use<br>• Adoption of industrial AI applications<br>• Job creation & hiring<br>• IP creation & licensing<br>• Energy consumption optimization<br>• Policy development & evaluation | • Aula & Bowles (2023)<br>• Birhane (2021)<br>• Cobbe, Veale, & Singh (2023)<br>• Gansky & McDonald (2022)<br>• Hind & Seitz (2022)<br>• Madianou (2021)<br>• Moore (2019)<br>• Widder & Nafus (2023) |

*High-level ethical concern 2: Issues arising from machine learning*

| Lower-level issues identified by Stahl et al. (2022) | Examples of related value chain actors | Examples of related resourcing activities | Related sources |
|---|---|---|---|
| *Control of data* (misuse of personal data, lack of privacy, security, integrity) | • Data subjects, owners, users, & brokers<br>• Data center operators<br>• Model developers<br>• Application users | • Privacy policy development, enforcement, & application to ML systems<br>• Informed consent and authorization for data collection, scraping, & use | • European Parliament (2020)<br>• MacKinnon & King (2022)<br>• Veale, Binns, & Edwards (2018)<br>• Crain (2018) |
| | • Cloud services providers<br>• Information security services providers<br>• Red teamers, blue teamers, & real adversaries<br>• Research funding sources & funding recipients | • Sale, purchase, brokerage, & ownership of data<br>• Knowledge of ML vulnerabilities & vulnerability testing methods<br>• Funding of ML research involving ethical concerns related to control of data | • Lamdam (2022)<br>• Fredrikson, Jha, & Ristenpart (2016)<br>• Greshake et al. (2023)<br>• Wang et al. (2022)<br>• Whittaker (2021)<br>• Hatmaker (2022)<br>• Perrigo (2022)<br>• Lee, Cooper, & Grimmelmann (2023)<br>• De Vynck (2023)<br>• GitHub Copilot Litigation (2023)<br>• Stable Diffusion Litigation (2023)<br>• Vincent (2023) |
| *Reliability* (Accuracy of predictive recommendations, accuracy of non-individualized recommendations, lack of quality data, accuracy of data) | • Data providers<br>• Model developers<br>• Cloudwork platform providers<br>• Application users & other application stakeholders<br>• Research funding sources & funding recipients | • Development and implementation of ethical quality assurance practices for model training, testing, & management<br>• Data annotation & verification and outsourcing of data work and model work<br>• Funding of ML research involving ethical concerns related to reliability | • Angwin et al. (2016)<br>• Bender et al. (2021)<br>• Burr & Leslie (2023)<br>• Eitel-Porter (2021)<br>• Grote & Berens (2022)<br>• Irani (2015)<br>• Miceli, Posada, & Yang (2022)<br>• Miceli & Posada (2022)<br>• Mökander & Axente (2023)<br>• Perrigo (2023)<br>• Rankin et al. (2020) |

| *Lack of transparency* (Bias and discrimination, lack of accountability and liability) | • Data subjects<br>• Model developers<br>• Application users & other application stakeholders<br>• Governments, courts, and regulatory bodies<br>• Research funding sources & funding recipients | • Incentivization & disclosure of funding for "responsible AI" and "AI ethics" research<br>• Documentation, disclosure, and explanation of machine learning & automated decision-making processes and outcomes<br>• Inclusion of stakeholder knowledge & perspectives<br>• Distribution and enforcement of accountability and liability for harms amongst value chain actors<br>• Community organizing & protest | • ACLU (2023)<br>• Ahmed, Wahed, & Thompson (2023)<br>• Bartneck et al. (2020)<br>• Birhane et al. (2022a, 2022b)<br>• Broderick (2023)<br>• Brown (2023)<br>• Cobbe, Veale, & Singh (2023)<br>• European Commission (2022)<br>• Miceli et al. (2022)<br>• Mitchell et al. (2019)<br>• Ochigame (2019)<br>• Raji et al. (2020)<br>• Whittaker (2021)<br>• Widder & Nafus (2023)<br>• Zech (2021) |
|---|---|---|---|

*High-level ethical concern 3: Ethics of living in a digital world*

| **Lower-level issues identified by Stahl et al. (2022)** | **Examples of related value chain actors** | **Examples of related resourcing activities** | **Related sources** |
|---|---|---|---|
| *Economic issues* (Unemployment, concentration of economic power, ownership of data/IP) | • Workers<br>• Unemployed & precariously employed people<br>• Employment services providers<br>• Education & training providers<br>• Employers | • Use of AI applications in hiring, contracting, dismissal, & surveillance of workers<br>• AI education & training<br>• Distribution & redistribution of capital, profits, and | • Ahmed, Wahed, & Thompson (2023)<br>• Bales & Stone (2020)<br>• Broderick (2023)<br>• Dastin (2018)<br>• Dyer-Witheford, |
| | • Tech companies<br>• Nations | other financial resources<br>• Unionization & collective organizing<br>• Distribution, open-sourcing, access to, & licensing of data, code, and other computational resources and IP | Kjøsen, & Steinhoff (2019)<br>• Hickok & Maslej (2023)<br>• Langenkamp & Yue (2022)<br>• Masiello & Slater (2023)<br>• Miceli, Posada, & Yang (2022)<br>• Miceli & Posada (2022)<br>• Perrigo (2023)<br>• Webster (2023)<br>• Widder, West, & Whittaker (2023) |
| *Justice* (Impact on justice systems, access to public services, impact on vulnerable groups, distribution, economic participation) | • Justice system administrators<br>• Public service administrators<br>• People seeking public services or justice<br>• Economically disadvantaged groups<br>• Exploited & precarious workers<br>• Employers | • Public and private funding of AI applications procured & used in justice systems and public services<br>• Data preparation, design, and development of AI systems with impacts on justice & subsequent impacts of decision automation outcomes for justice<br>• Inclusion of knowledges from vulnerable and marginalized groups in AI education, design, development, & governance processes<br>• Redistribution of resources required to justly develop/use | • Angwin et al. (2017)<br>• Birhane et al. (2022a)<br>• Dyer-Witheford, Kjøsen, & Steinhoff (2019)<br>• Eubanks (2018)<br>• Gans-Combe (2022)<br>• Mulligan & Bamberger (2019)<br>• Pasquale (2020)<br>• Solaiman et al. (2023)<br>• West, Whittaker, & Crawford (2019) |
| | | AI systems & just distribution of value co-created throughout AI system lifecycles<br>• Macro-level social, political, and economic outcomes of widespread AI adoption | |
| *Human freedoms* (Loss of freedom and individual autonomy, harm to physical integrity, impact on health, lack of access to and freedom of information, reduction of human contact) | • Exploited workers<br>• Victims of material, psychological, & environmental harms<br>• Government & corporate information owners<br>• Companies, governments, researchers, and civil society actors seeking access to data and information | • Compensation for labor and resulting gains/losses to social, political, & economic freedoms and autonomy<br>• Algorithmic discrimination resulting in reduction of social, political, & economic opportunities<br>• Energy usage of AI models resulting in carbon emissions, water usage of data centers resulting in depletion of freshwater reserves<br>• Creation, collection, and brokerage of access to proprietary data, information, and computational resources | • Ahmed, Wahed, & Thompson (2023)<br>• Angwin et al. (2016)<br>• Eubanks (2018)<br>• GPAI (2022)<br>• Miceli & Posada (2022)<br>• Whittaker (2021) |
| *Broader societal issues* (Potential for military use, impact on environment, impact on democracy) | • Governments & police services<br>• Military organizations & defense contractors<br>• Cloud services providers<br>• Data center operators<br>• Hardware manufacturers<br>• Social media platform | • Military and police procurement, contracting, and use of AI applications<br>• Mining activities & mineral extraction<br>• Fuel required to ship & transport materials & hardware<br>• Hardware manufacturing & assembly<br>• Energy, mineral, & water consumption of AI systems | • Crawford (2021)<br>• Crawford & Joler (2018)<br>• GPAI (2021)<br>• Hoijtink & Hardeveld (2022)<br>• Krönke (2019)<br>• Li et al. (2023)<br>• Luccioni & Hernandez- |
| | providers & users | • Disposal & recycling of e-waste at end of hardware lifecycle<br>• Creation & reinforcement of filter bubbles based on algorithmic profiling of users | Garcia (2023)<br>• Mahoney (2020)<br>• Mulligan & Bamberger (2019)<br>• Taddeo et al. (2021)<br>• Woolley (2018) |
| *Unknown issues* (Unintended or unforeseeable adverse impacts, cost to innovation, potential for criminal and malicious use, prioritization of the "wrong" problems) | • Malicious actors<br>• Security professionals<br>• Governments & regulatory bodies<br>• Research funders, research institutes, & researchers | • Unforeseen misuses/abuses of personal data & digital identities<br>• Unforeseen consequences of social engineering, identity theft, ML model hacking, mis/disinformation, and malicious data brokerage operations<br>• Implementation & enforcement of excessively strict or excessively permissive AI regulation<br>• Excessive funding of misdirected AI research projects | • Ada Lovelace Institute (2021)<br>• Brundage et al. (2018)<br>• Smuha (2021)<br>• Tiku (2023) |

*High-level ethical concern 4: Metaphysical issues*

| **Lower-level issues identified by Stahl et al. (2022)** | **Examples of related value chain actors** | **Examples of related resourcing activities** | **Related sources** |
|---|---|---|---|
| Machine consciousness, autonomous moral agents, super-intelligence, singularity, changes to human nature | • Governments<br>• Intergovernmental bodies<br>• Individual humans<br>• Human organizations<br>• Hypothetical conscious machines<br>• Hypothetical artificial moral agents<br>• Hypothetical superintelligent agents | • Data & knowledge assembled to develop hypothetical conscious machines & superintelligent agents<br>• Consolidation of resources under the control of hypothetical conscious machines & superintelligent agents<br>• Uneven distribution of hypothetical AI & technological capabilities resulting in divergent evolutionary trajectories | • Gebru & Torres (2023)<br>• Torres (2023) |